\documentclass[aps,prd,superscriptaddress,showpacs,twocolumn,nofootinbib]{revtex4-1}
\usepackage{graphicx}
\usepackage[caption=false]{subfig}
\usepackage{epstopdf}
\usepackage{amsmath}
\usepackage{amsfonts} 
\usepackage{amssymb}
\usepackage{latexsym}
\usepackage{hyperref}
\usepackage[english]{babel}
\usepackage[utf8]{inputenc}
\usepackage[dvipsnames]{xcolor}
\usepackage{slashed}
\usepackage{feynmp}
\usepackage{bm}
\usepackage{bbold}
\usepackage{eufrak}
\usepackage{tabu}

\begin{document}

\title{Constraining the photon mass via Schumann resonances}

\author{P. C. Malta}\email{pedrocmalta@gmail.com}
\affiliation{R. Antonio Vieira 23, 22010-100, Rio de Janeiro, Brazil}

\author{J. A. Helay\"{e}l-Neto}\email{helayel@cbpf.br}
\affiliation{Centro Brasileiro de Pesquisas F\'{i}sicas (CBPF), Rua Dr Xavier Sigaud 150, Urca, Rio de Janeiro, Brazil, CEP 22290-180}

%%%%%%%%%%%%%%%%%%%%%%%%%%%%%%%%%

\begin{abstract}

The photon is the paradigm for a massless particle and current experimental tests set severe upper bounds on its mass. Probing such a small mass, or equivalently large Compton wavelength, is challenging at laboratory scales, but planetary or astrophysical phenomena may potentially reach much better sensitivities. In this work we consider the effect of a finite photon mass on Schumann resonances in the Earth-ionosphere cavity, since the TM modes circulating Earth have eigen-frequencies of order $\mathcal{O} (10 \, {\rm Hz})$   that could be sensitive to $m_\gamma \approx 10^{-14} \, {\rm eV/c}^2$. In particular, we update the limit from Kroll [Phys. Rev. Lett. {\bf 27}, 340 (1971)], $m_\gamma \leq 2.4 \times 10^{-13} \, {\rm eV/c}^2$, by considering realistic conductivity profiles for the atmosphere. We find the conservative upper bound $m_\gamma \leq 2.5 \times 10^{-14} \, {\rm eV/c}^2$, a factor 9.6 more strict than Kroll's earlier projection.

\end{abstract}

%\pacs{11.10.Ef, 11.15.Wx, 11.15.Bt.}
\maketitle

%%%%%%%%%%%%%%%%%%%%%%%%%%%%%%%%%%%

\section{Introduction}  \label{sec_intro}
\indent

At the end of the 19th century Maxwell unified electricity and magnetism and realized that electromagnetic waves propagate at a fixed speed determined by the properties of the vacuum, $c = 1/\sqrt{\varepsilon_0 \mu_0}$. Hertz proved that light moves at this speed, thereby showing that light is an electromagnetic wave with energy $E$ carrying linear momentum $p = E/c$. In Einstein's 1905 {\it annus mirabilis} he showed, among other things, that the photoelectric effect could be explained if light would also behave as a particle. He also demonstrated that mass and energy are related via the dispersion relation $E = \sqrt{m^2 c^4 + p^2 c^2}$, where $m$ is the particle's rest mass. Thus, these results indicate that light is a particle -- the photon -- and its rest mass, $m_\gamma$, must be identically zero.

This prediction is of fundamental consequence and may lend itself to experimental verification. The most obvious consequence of a finite photon mass is a change in the dispersion relation of light causing violet and red radiation to move at different speeds, an effect that could be tested with astrophysical observations. Field configurations are also modified: a point electric charge produces a screened Yukawa -- rather than Coulomb -- potential with a screening scale $\lambda_\gamma \sim m_\gamma^{-1}$, which is also the photon's Compton wavelength. Given the purported smallness of the photon mass, $\lambda_\gamma$ is expected to be very large, so only large distance scales -- or long time periods -- are relevant. Therefore, the most promising way to probe a finite photon mass is to use long-range, quasi-static electromagnetic phenomena.

Recent limits on the photon mass are listed in Ref.~\cite{PDG}. The tightest limit, $m_\gamma \leq 10^{-18} \, {\rm eV/c^2}$, was obtained using solar wind data from the Voyager missions at Pluto's orbit (40~AU)~\cite{MHD}. Other strong upper bounds were extracted through the analysis of fast radio bursts~\cite{Bonetti, Bentum, Wang}, solar wind data at 1~AU~\cite{Retino}, Jovian magnetic-field measurements~\cite{Davis} and null tests of Coulomb's law~\cite{Williams}. For comprehensive reviews, see refs.~\cite{Tu, Nieto, Okun, Goldhaber}. As previously indicated, the strongest limits required either exquisitely precise or large-scale experiments, a general tendency when constraining a finite photon mass~\cite{Goldhaber, Goldhaber2}.

Measurements of terrestrial phenomena have also been used to establish robust upper bounds. Fischbach {\it et al.} found $m_\gamma \leq 8 \times 10^{-16} \, {\rm eV/c^2}$ by studying geomagnetic fields in light of the modified Amp\`ere's law~\cite{Fischbach}. F\"ullekrug used the variations in the speed of radio waves in the terrestrial atmosphere due to changes in the reflection height to obtain $m_\gamma \leq 2 \times 10^{-16} \, {\rm eV/c^2}$~\cite{Fuellekrug}, though this result has been criticized~\cite{Goldhaber}. Finally, Kroll studied Schumann resonances on Earth to obtain $m_\gamma \leq 2.4 \times 10^{-13} \, {\rm eV/c^2}$~\cite{Kroll1, Kroll2}. Let us discuss this last result in more detail.

Since the 1890s it has been conjectured that electric excitations in the atmosphere would produce resonating waves parallel to and between the conducting surface at $r = R_\oplus \approx 6371$~km and the lower layers of the ionosphere at heights $z \approx 100$~km (D region). Inside a conductor the electric field is zero and its tangential component is continuous across boundaries. Keeping in mind that, in the context of a spherical waveguide, transversality is defined relative to the radial direction, transverse electric (TE) modes must have a variation of at least half a wavelength to fulfil the boundary conditions at $R_\oplus$ and $R_\oplus + z$, meaning that the resonant frequencies are $f_{\rm TE} \sim c/z \sim$~kHz. Transverse magnetic (TM) modes, on the other hand, have electric fields satisfying the boundary conditions with much less variation, so that $f_{\rm TM} \sim c/R_\oplus \sim$~Hz. In fact, for an empty cavity with $z \ll R_\oplus$, the eigen-frequencies are 
\begin{equation} \label{eq_SR_ideal}
f_{\ell}= \frac{c}{2\pi R_\oplus} \sqrt{ \ell(\ell + 1) } \, ,
\end{equation}
giving 10.6, 18.4 and 25.9~Hz for $\ell = 1,2,3$, respectively. These are the so-called Schumann frequencies~\cite{schumann_orig}, though W.O. Schumann was not the first to obtain this result~\cite{JacksonHistory, Besser2007}.

These extremely-low frequency (ELF) waves were measured by Balser and Wagner in 1960 and the frequencies of the first three modes were found to be 7.8, 14.1 and 20.3~Hz~\cite{Balser}, {\it i.e.}, $\sim 20\%$ lower than those predicted by Eq.~\eqref{eq_SR_ideal}. This is due to the fact that neither Earth's surface nor the atmosphere are perfect conductors, meaning that the quality factor of the cavity is finite, thus shifting the resonant frequencies downwards~\cite{Jackson}. Furthermore, the cavity is not empty, but filled with air possessing a finite conductivity profile. This last remark is fundamental, since the details of the profile heavily influence the propagation of ELF waves in the atmosphere.

The study of Schumann resonances offers interesting applications. The most common sources are large electric transients, such as cloud-to-ground lightning~\cite{Pfaff}. It has been suggested to track worldwide lightning activity through precise measurements of the ELF spectrum, allowing the inference of temperature fluctuations in the atmosphere. Schumann resonances could then act as a global thermometer~\cite{Williams1992, Hobara}, as well as a monitor of the tropospheric water vapor concentration~\cite{Price2000}. It has also been suggested that earthquakes could be forecast by searching for pre-seismic perturbations in the ELF spectrum caused by ionospheric depressions around the epicenter~\cite{quakes}. Disturbances in the ELF spectrum have also been observed after the Johnston Island high-altitude nuclear test of July 9, 1962 (``Starfish Prime" test at an altitude of 400~km)~\cite{Madden}. Also noteworthy are possible effects of ELF waves on human health~\cite{health_0, health_1, health_2}.

Let us now return to Kroll's works. In Ref.~\cite{Kroll1} waveguides and resonant cavities are discussed in the context of a massive photon, showing that the empty-space dispersion relation of a massive photon\footnote{The frequency of the photon is $k c$ and $\hbar \kappa/c$ is its rest mass.}, $k'^2 = k^2 + \kappa^2$, is not generally valid, though this relation is approximately correct for a spherical cavity. However, this is no longer the case in a cavity composed of two conducting spherical shells~\cite{Kroll2}. Consequently, he writes $k'^2 = k^2 + g\kappa^2$, with $g$ being a mass sensitivity coefficient depending on the radii of the shells and $k$, and proceeds to obtain the limit $\lambda_\gamma/2\pi \geq 8.3 \times 10^7$~cm, or $m_\gamma \leq 4.8 \times 10^{-46}$~g.

It is important, however, to mention a few caveats of his approach. Even though he works out the boundary conditions for the now physically meaningful scalar and vector potentials for the case of finite conductivity, his limit does not take relevant features of the Earth-ionosphere cavity into consideration, namely finite conductivities at the boundaries and a conductivity profile for the atmosphere. In fact, he explicitly assumes perfectly conducting shells and a nominal height of 70~km for the (empty) ionosphere. In his own words, the author confines himself ``to a crude approximation", where he uses the mass sensitivity coefficient $g$ obtained in the limit of infinite conductivity. It is the goal of this paper to improve Kroll's limit by taking these important points into account.

This paper is organized as follows: in Sec.~\ref{sec_theory} we discuss the de Broglie-Proca theory in a conducting medium. In Sec.~\ref{sec_results} we present realistic conductivity profiles, extracting the eigen-frequencies and quality factors as a function of the photon mass. Comparing these results with observations, we set upper bounds on the photon mass. Our concluding remarks are presented in Sec.~\ref{sec_conclusions}. We use SI units and spherical coordinates $\left( r, \theta, \varphi \right)$ throughout.

 %%%%%%%%%%%%%%%%%%%%%%%%%%%%%%%%%%%%%%%%%%

\section{Theoretical setup}  \label{sec_theory}
\indent

The photon, $A^\mu = \left( \phi/c, {\bf A} \right)$, now with mass $m_\gamma$, is described by the de Broglie-Proca Lagrangian~\cite{dB1, dB2, dB3, Proca1, Proca2}
\begin{equation} \label{lag_proca}
\mathcal{L} = -\frac{1}{4\mu_0} F^{\mu\nu} F_{\mu\nu} + \frac{\mu_\gamma^2}{2\mu_0} A_\mu A^\mu   - J_\mu A^\mu \, ,
\end{equation}
where $J^\mu = \left( c\rho, {\bf J} \right)$ is the 4-current density. The anti-symmetric field-strength tensor is $F_{\mu\nu} = \partial_\mu A_\nu - \partial_\nu A_\mu$ with the electric and magnetic fields given by $F_{0i} = {\bf E}_i/c$ and $F_{ij} = -\varepsilon_{ijk}{\bf B}_k$, respectively. The fields are defined in terms of the potentials as usual (${\bf B} = \mu_0 {\bf H}$)
\begin{equation} \label{eq_def_H_E}
{\bf H} = \frac{1}{\mu_0} \nabla\times {\bf A} \quad\quad {\rm and} \quad\quad {\bf E} = -\nabla\phi + i\omega {\bf A} \, .
\end{equation}
From Eq.~\eqref{lag_proca} we obtain the de Broglie-Proca equation
\begin{equation} \label{eq_proca}
\partial_\mu F^{\mu\nu} + \mu_\gamma^2 A^\nu = \mu_0  J^\nu \, 
\end{equation}
and the constraint $\partial_\mu A^\mu = 0$ is automatically enforced if local charge conservation, $\partial_\mu J^\mu = 0$, is valid. Note that this is a subsidiary condition, not a gauge choice, and the lack of gauge symmetry of Eq.~\eqref{eq_proca} implies that both potentials and field strengths are physically meaningful.

Here $\mu_\gamma = m_\gamma c/\hbar$ is the reciprocal (reduced) Compton wavelength and may be conveniently expressed as 
\begin{equation} \label{eq_mu}
\mu_\gamma  = \frac{0.3}{R_\oplus} \left( \frac{m_\gamma}{10^{-14} \, {\rm eV/c^2}} \right) 
\end{equation}
with Earth's mean radius $R_\oplus \approx 6371$~km. This indicates that experiments and phenomena at planetary scales will be sensitive to photon masses $m_\gamma \sim 10^{-14} \, {\rm eV/c^2}$.

The current density is ${\bf J} = {\bf J}_{\rm con} + {\bf J}_{\rm ext}$. The first term represents the current due to the local atmospheric conductivity, given by Ohm's law: ${\bf J}_{\rm con} = \sigma {\bf E}$. The second describes external sources, but our main focus here is to determine the (generally complex) frequencies of the normal modes, so we set ${\bf J}_{\rm ext} = 0$~\cite{Jackson, Sentman}. The main sources are lightning events, which incoherently excite the Earth-ionosphere cavity roughly 40 times per second (global average)~\cite{Oliver_lightning}, with flashes lasting $\lesssim 0.5$~s (median)~\cite{Kakona, Lopez}. Knowledge of the external sources ({\it e.g.}, currennt spectrum and location) is nonetheless required to realistically assess field amplitudes and spectra at a receiver~\cite{Sentman96} (see also Sec.~\ref{sec_results}).

%The current density is ${\bf J} = {\bf J}_{\rm con} + {\bf J}_{\rm ext}$. The first term represents the current due to the local atmospheric conductivity, given by Ohm's law: ${\bf J}_{\rm con} = \sigma {\bf E}$. The second term \textcolor{red}{describes lightning, the main external sources, that incoherently excite the cavity roughly 40 times per second (global average)~\cite{Oliver_lightning}, with each flash lasting $\lesssim 0.5$~s (median)~\cite{Kakona, Lopez}. In what follows we solve the de Broglie-Proca equations after the injection of radiation into the cavity by lightning, that is, ${\bf J}_{\rm ext} = 0$, allowing us to focus on the eigen-frequencies and quality factors. Knowledge of the external sources ({\it e.g.}, currennt spectrum and location) is nonetheless required to realistically determine field amplitudes and spectra~\cite{Sentman96}, but this is beyond the scope of this work (see also).     }

Returning to the de Broglie-Proca equations, let us assume a harmonic $e^{-i\omega t}$ time dependence for fields and potentials. With this, Eq.~\eqref{eq_proca}, together with the usual Bianchi identities, becomes
\begin{subequations}
\begin{eqnarray}
\nabla\cdot {\bf E} & = & \frac{\rho}{\varepsilon_0}  - \mu_\gamma^2 \phi \, , \label{eq_gauss2} \\ 
\nabla\cdot {\bf H} & = & 0 \, , \label{eq_mono2} \\ 
\nabla\times {\bf E} & = & i\mu_0\omega {\bf H} \, , \label{eq_induct2} \\ 
\nabla\times {\bf H} & = & - i\varepsilon_0 \omega n^2 {\bf E} - \frac{\mu_\gamma^2}{\mu_0} {\bf A} \, , \label{eq_ampere2}
\end{eqnarray}
\end{subequations}
where the position-dependent refraction index (squared) is given by
\begin{equation} \label{eq_n2}
n^2(r) = 1 + \frac{i\sigma(r)}{\varepsilon_0 \omega} \, .
\end{equation}

Finally, it is necessary to state the appropriate boundary conditions for the de Broglie-Proca electrodynamics. As discussed in Ref.~\cite{Kroll1}, the scalar and vector potentials are continuous everywhere, thus implying that the electric and magnetic fields are subject to the same boundary conditions as in Maxwell's electrodynamics~\cite{Jackson}, independently of the photon mass. Furthermore, the energy input ({\it e.g.}, lightning) lies within the bulk of the cavity and is dissipated outwards, requiring that adequate conditions be imposed on outgoing waves. Let us now turn our attention to the regions of interest, namely the interior of the Earth and the atmosphere.

%%%%%%%%%%%%%%%%%%%
\subsection{Earth's interior ($r \leq R_\oplus$)} \label{sec_const_cond}
\indent

The terrestrial surface represents the lower boundary of the resonating Earth-ionosphere cavity. Measured values for the conductivity of the crust (depth $\lesssim 30$~km) at the ELF range vary considerably due to different ground composition with $\sigma \approx 10^{-4}-10^{-2}$~S/m, whereas $\sigma \approx 4$~S/m for seawater at depths $\lesssim 10$~km~\cite{world_atlas_cond, Maus}. The net negative charge at the terrestrial surface is $\approx 10^6$~C~\cite{Rycroft}. Also the upper and lower mantles at depths in the range $\sim 30-1000$~km have relatively high conductivities: $\sigma \approx 10^{-2} - 10$~S/m~\cite{Price1, Poirier, mantle2}. These values are much larger than those found in the atmosphere, particularly near the ground, cf. Sec.~\ref{sec_inhom_cond}.

As already stated, the boundary conditions for the electric and magnetic fields in massive electrodynamics are the same as in the massless case. It is thus interesting to determine the respective equations of motion inside Earth, where we assume a very large and, for all practical purposes, constant conductivity. Taking the curl of Eq.~\eqref{eq_ampere2} and plugging Eq.~\eqref{eq_induct2}, we find
\begin{equation} \label{eq_wave_H}
\nabla^2 {\bf H}  + \left( k^2 n^2 - \mu_\gamma^2  \right) {\bf H}  = 0 \, ,
\end{equation}
where $k = \omega/c$. The electric field satisfies
\begin{equation} \label{eq_wave_E}
\nabla^2{\bf E} + \left( k^2 n^2 - \mu_\gamma^2 \right){\bf E} = -(n^2 - 1) \nabla \left( \nabla\cdot{\bf E}  \right)  \, .
\end{equation}
%
%\textcolor{blue}{and both equations display a diffusive behavior of the fields in a region of very high conductivity}.

Most relevant to our present discussion is the observation that, in regions of high conductivity (or formally $|n^2| \rightarrow \infty$), Eqs.~\eqref{eq_wave_H} and~\eqref{eq_wave_E} indicate that both electric and magnetic fields vanish, independently of $m_\gamma$~\cite{Kroll2}. The situation is analogous to that of Maxwell's electrodynamics, in which the electromagnetic fields are zero inside a perfect conductor. This fact will be useful in Sec.~\ref{sec_results}, when we set the boundary conditions at $r = R_\oplus$. Note, however, that this conclusion is not valid for the vector and scalar potentials, which are finite and carry energy within Earth's interior~\cite{Kroll2, Tu}.

Given that $\sigma \gtrsim 10^{-3}$~S/m for $r \leq R_\oplus$, let us investigate how deep the fields penetrate Earth in the massive case. Naively assigning $\nabla \rightarrow i\beta$ to Eq.~\eqref{eq_wave_H}, we get
\begin{equation}
\beta^2 = \left( \omega^2/c^2 - \mu^2_\gamma \right) + i\mu_0\omega \sigma \, ,
\end{equation}
whose square root is $\beta = \beta_{+} + i \beta_{-}$ with
\begin{equation} \label{eq_skin}
\beta_{\pm} \! = \! \frac{\omega}{c \sqrt{2}} \sqrt{ \sqrt{\left( 1 - \frac{c^2\mu_\gamma^2}{\omega^2} \right)^2 + \left( \frac{\sigma}{\varepsilon_0 \omega} \right)^2}  \pm \left( 1 - \frac{c^2\mu_\gamma^2}{\omega^2} \right)   }  \, .
\end{equation}
This rough estimate does not take the exact geometry of the problem into consideration. Nonetheless, it clearly shows that the magnetic field displays a diffusive behavior in a region of very high conductivity, being damped within the conducting medium with a characteristic length $L = 1/\beta_{-}$, the so-called skin depth~\cite{Jackson}. The electric field is similarly damped, also exhibiting a diffusive character.

\begin{figure}[t!]
\begin{minipage}[b]{1.0\linewidth}
\includegraphics[width=\textwidth]{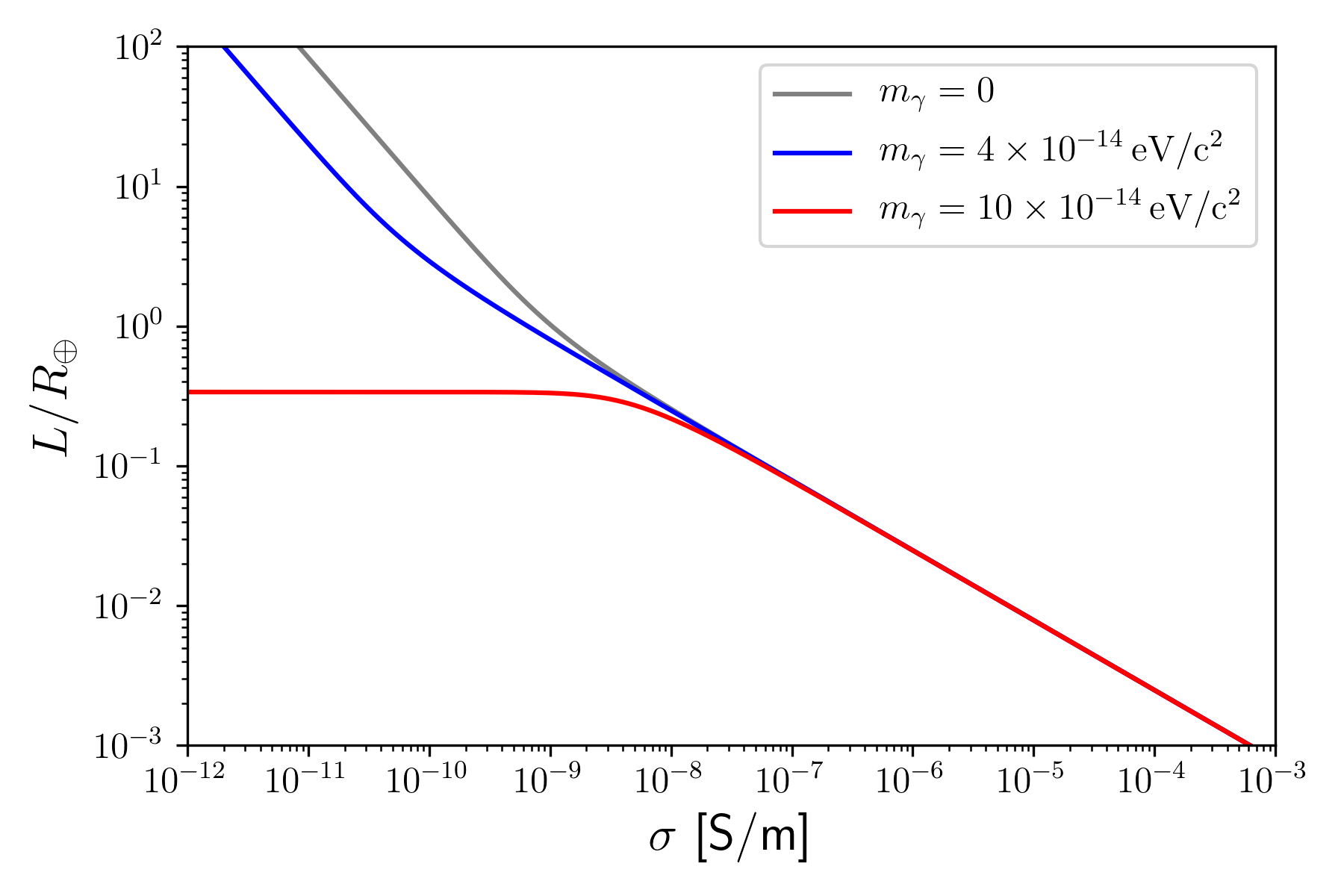}
\end{minipage} \hfill
\caption{Skin depth in units of Earth's radius as a function of conductivity, cf. Eq.~\eqref{eq_skin}, with $f = 10$~Hz.}
\label{fig_skin_vs_sigma}
\end{figure}

As in the massless Maxwell case, the penetration length depends on the frequency of the impinging radiation and on the conductivity of the medium. In the present case, however, the photon mass also plays a role through the dimensionless ratio
\begin{equation}
\left( \frac{c \mu_\gamma}{\omega} \right)^2 = 0.059 \left( \frac{10 \, {\rm Hz}}{f} \right)^2 \left( \frac{m_\gamma}{10^{-14} \, {\rm eV/c^2}} \right)^2 \, .
\end{equation}
It is clear that effects of a finite photon mass are only relevant for low enough conductivities. For large photon masses ($\mu_\gamma \gg \omega/c$), if $\sigma \ll \mu_\gamma^2/\mu_0\omega$, the skin depth becomes independent of frequency and conductivity, being given by $L \approx \sqrt{2}/\mu_\gamma$. This is illustrated in Fig.~\ref{fig_skin_vs_sigma}, as well as the general behavior for different values of the photon mass.

At ELF and with conductivities in the range characteristic of our problem, cf. Fig.~\ref{fig_cond_profiles}, if $m_\gamma \lesssim 10^{-13} \, {\rm eV/c^2}$, the usual result from Maxwell's electrodynamics remains a good approximation. Furthermore, for the values quoted above for the Earth ($r \leq R_\oplus$) we have $L \lesssim \mathcal{O}(10 \, {\rm km}) \ll R_\oplus$, cf. Fig.~\ref{fig_skin_vs_sigma}, and we are therefore able to assume Earth to be a perfect conductor, in particular when compared to the lower atmosphere, cf. Sec.~\ref{sec_inhom_cond}. In fact, even if we use the actual, finite conductivity of Earth's surface, we expect the results to be essentially independent of it~\cite{cole}.

%%%%%%%%%%%%%%%%%%%
\subsection{Atmosphere ($r > R_\oplus$)} \label{sec_inhom_cond}
\indent

The most relevant feature of the atmosphere is its electric conductivity. Unfortunately, direct experimental data are scarce. Aircraft measurements can be made only up to 15~km or with meteorological balloons up to 35~km; between 35 and 100~km only geophysical rockets may be used~\cite{ACDC}. Thus, one may not rely entirely on experimental input and one typically solves the inverse problem: given the measured Schumann spectrum, theoretical modelling is used to validate tentative conductivity profiles. If the projected properties (such as frequencies, quality factors, etc) agree well with observations, the profile is validated.

Earth's atmosphere may be roughly divided in two regions. The lower region is dominated by ions and has a conductivity $\sigma \approx 10^{-13}$~S/m due to ground radioactivity. The conductivity rapidly increases with height and the upper layer is dominated by free electrons due to solar and cosmic irradiation~\cite{Burke}. The transition from ion- to electron-dominated regions happens at $\approx 60-70$~km at the so-called conductivity height where $\sigma \approx \varepsilon_0 \omega$ -- here radiation moves from a wave-like to a diffusion-like behavior. At heights $\sim 100$~km the conductivity varies from $\sigma \approx 10^{-6}$~S/m at night to $\sigma \approx 10^{-3}$~S/m during the day and radio waves are effectively reflected~\cite{Greifinger}.

\begin{figure}[t!]
\begin{minipage}[b]{1.02\linewidth}
\includegraphics[width=\textwidth]{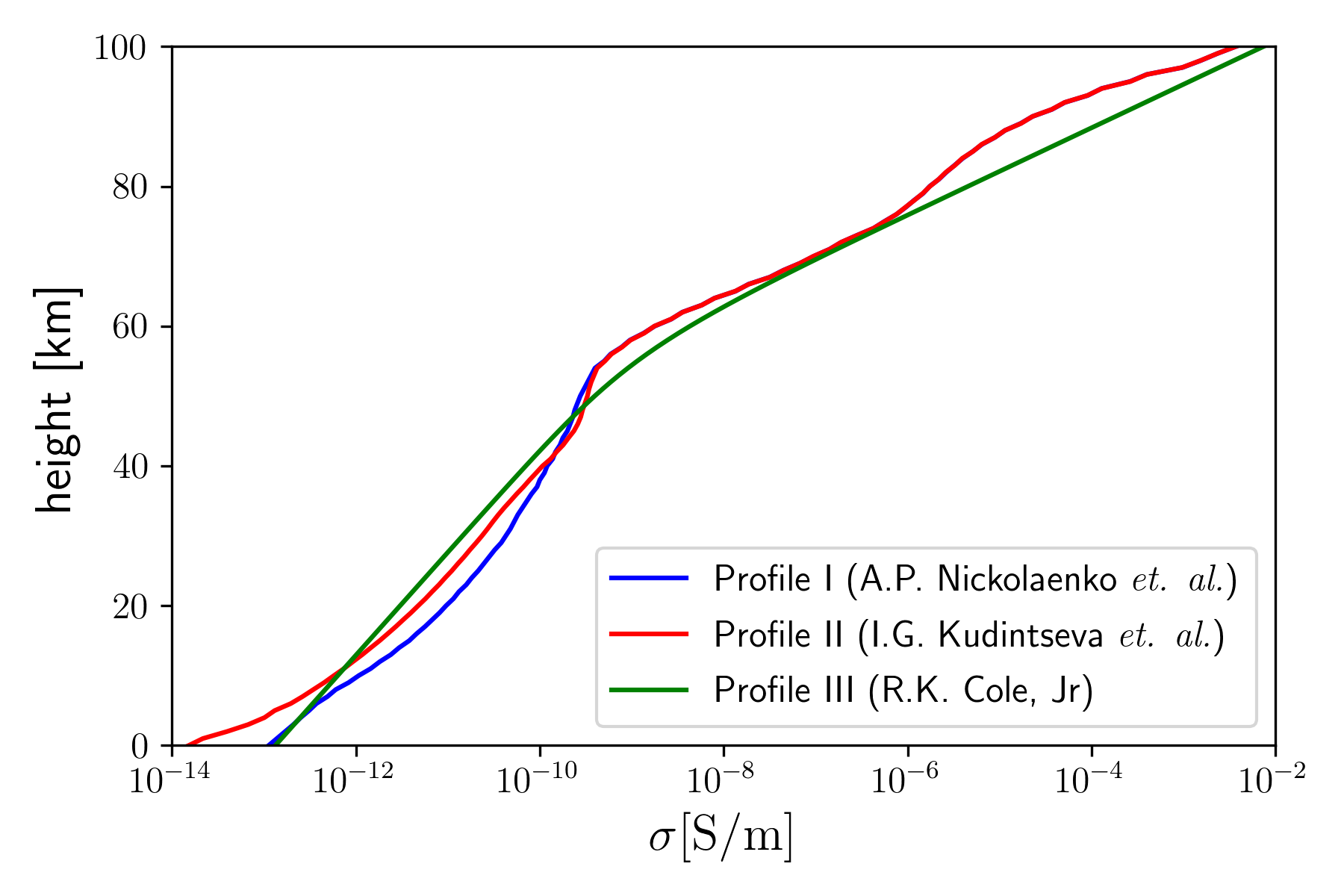}
\end{minipage} \hfill
\caption{Conductivity profiles. Profile~I~\cite{cond_prof_galuk} ranges from $z = 0-99$~km; profile~II~\cite{ACDC} goes from $z = 2-98$~km. Both were extended to $z = 0 - 100$~km by linearly extrapolating $\log_{10} \sigma$. Profile III is from Cole~\cite{cole}.}
\label{fig_cond_profiles}
\end{figure}

In what follows we shall ignore such day-night asymmetries (and also those from the geomagnetic field)  and model the conductivity of the atmosphere through isotropic, spherically stratified profiles, {\it i.e.}, as scalar functions of the altitude, $\sigma = \sigma(z)$, with $z = r - R_\oplus$, since such profiles fit measured data well. For the sake of concreteness, we consider the recent numerical estimates of the conductivity profiles from refs.~\cite{ACDC, cond_prof_galuk}, as well as the analytical model from Cole (profile~III in Ref.~\cite{cole}). The chosen profiles, illustrated in Fig.~\ref{fig_cond_profiles}, support the features discussed above and display the well-known ``knee" at $\approx 60$~km.

The isotropic and inhomogeneous atmosphere supports the propagation of TM modes~\cite{Bliokh} and the radial variation of the index of refraction will play a crucial role. The wave equation for the vector potential is
\begin{equation}  
\nabla^2{\bf A} + \left( k^2 - \mu_\gamma^2 \right){\bf A} = -\mu_0 {\bf J} \, ,
\end{equation}
but, using Ohm's law, we may rewrite it as~\cite{Sentman, Titan} 
\begin{equation} \label{eq_wave_A_phi}
\nabla \times \nabla \times {\bf A} = i \frac{k n^2}{c} \nabla\phi + \left( k^2 n^2 - \mu_\gamma^2 \right) {\bf A} \, .
\end{equation}
As it stands, this equation is also valid in Maxwell's theory by setting $\mu_\gamma = 0$.

We are interested in the Schumann resonances, {\it i.e.}, the cavity modes with lowest eigen-frequencies. Given that the boundary conditions satisfied by the electric and magnetic fields are the same as in massless electrodynamics, these ELF waves will also correspond to TM modes, for which $H_r = 0$. Contrary to refs.~\cite{Kroll1, Kroll2}, we retain generality and allow $\varphi$-dependent fields and potentials. Thus, from Eq.~\eqref{eq_def_H_E} we have
\begin{equation}
\mu_0 H_r = \frac{1}{r \sin\theta} \left[   \frac{\partial \left( \sin\theta A_\varphi \right)}{\partial\theta} - \frac{\partial A_\theta}{\partial\varphi} \right] = 0 \, ,
\end{equation}
which must be valid for all $r, \theta, \varphi$. For this to be true in general, we require that $A_\theta (r, \theta , \varphi) = A_\varphi (r, \theta, \varphi) = 0$, so that ${\bf A} (r, \theta, \varphi)  = A_r (r, \theta, \varphi) \hat{{\bf e}}_r$.

A single scalar function, $A_r(r, \theta, \varphi)$, controls the electrodynamics, acting as a Hertz potential~\cite{Jackson, Bliokh}. Since the vector potential points along the radial direction, the following identity holds
\begin{eqnarray}
\nabla\times\nabla\times{\bf A} &=& \frac{\ell (\ell + 1)}{r^2} A_r(r) Y_{\ell m}(\theta, \varphi) \hat{{\bf e}}_r   \\
&+&  \frac{1}{r}\frac{d A_r (r)}{dr} \left( \frac{\partial Y_{\ell m}}{\partial\theta} \hat{{\bf e}}_\theta + \frac{1}{\sin\theta} \frac{\partial Y_{\ell m}}{\partial\varphi} \hat{{\bf e}}_\varphi  \right) \, . \nonumber
\end{eqnarray}
Here radial and angular variables were separated as usual, $A_r(r, \theta, \varphi) = A_r(r) Y_{\ell m}(\theta, \varphi)$, with $Y_{\ell m}(\theta, \varphi)$ being the standard spherical harmonics~\cite{Arfken}. Moreover, because of the subsidiary condition, the scalar potential may be similarly split into $\phi(r, \theta, \varphi) = \phi_r(r) Y_{\ell m}(\theta, \varphi)$.

With this, from Eq.~\eqref{eq_wave_A_phi} we obtain the so-called height-gain functions~\cite{Titan}
\begin{equation}
\frac{dA_r}{dr} = i\frac{k n^2}{c} \phi_r \! \quad {\rm and} \quad \! \frac{d\phi_r}{dr} = ick \left( 1 - \frac{\gamma^2}{k^2 n^2} \right) A_r ,
\end{equation}
where we defined 
\begin{equation}
\gamma^2 = \mu_\gamma^2 + \frac{\ell(\ell + 1)}{r^2} \, .
\end{equation}
Decoupling the system above, we get
\begin{eqnarray}
\frac{d^2 A_r}{dr^2} + (n^2\nabla n^{-2}) \frac{dA_r}{dr} + \left( k^2 n^2 - \gamma^2 \right)A_r = 0  \,  \label{eq_dyn_Ar}
\end{eqnarray}
with a similar equation for $\phi_r (r)$, which is omitted.

As a closing comment we would like to mention that, besides the locally varying conductivity, also the radial profile of the electric permitivity could have been taken into account by making $\varepsilon_0 \rightarrow \varepsilon_0 \varepsilon_r(r)$. In the case of Schumann resonances on Earth we are allowed to ignore any spatial variations, as the pressures and temperatures involved are relatively low and do not significantly impact ELF waves. Incidentally, this is not a good approximation for ELF waves in other planets such as Venus~\cite{Hamelin}.

%%%%%%%%%%%%%%%%%%%%%%%%%%%%%%%%%%%%%%%%%%

\section{Analysis}  \label{sec_results}
\indent

Equation~\eqref{eq_dyn_Ar} is identical in Maxwell's theory provided $k^2 n^2 \rightarrow k^2 n^2 - \mu_\gamma^2$~\cite{Titan}. This similarity allows us to follow the method outlined in Ref.~\cite{Bliokh} and conveniently re-write $A_r(r, \theta, \varphi)$ in terms of a new scalar function (Hertz potential) $U(r, \theta, \varphi)$ as 
\begin{equation} \label{eq_def_U}
A_r(r, \theta, \varphi) = - \frac{i\omega \sqrt{n^2}}{c^2} r U(r, \theta, \varphi) \, .
\end{equation}
Since $A_r(r, \theta, \varphi) \sim Y_{\ell m}(\theta, \varphi)$, we have $U(r,\theta,\varphi) = u_\ell (r) Y_{\ell m} (\theta, \varphi)$, so that Eq.~\eqref{eq_dyn_Ar} becomes
\begin{equation} \label{eq_Ar_full}
\left[ \frac{d^2}{dr^2}  + (k^2 n^2 - \gamma^2) - \sqrt{n^2} \frac{d^2}{dr^2} \left( \frac{1}{\sqrt{n^2}} \right) \right]  \left( r u_\ell (r) \right) = 0 \, . 
\end{equation}
This equation could be used to extract the Schumann spectrum within a so-called full-wave treatment~\cite{Wait1970}, where the atmosphere is sliced in thin spherical shells within which the conductivity is approximately constant. Equation~\eqref{eq_Ar_full} is then solved within each slab using the adequate boundary conditions, thus producing a system of coupled algebraic equations for the amplitudes of the vector potential. Here we follow an alternative approach.

Instead of solving Eq.~\eqref{eq_Ar_full} in terms of the less familiar vector potential via the full-wave method, let us consider the normalized spherical impedance defined as~\cite{Bliokh}
\begin{equation} \label{eq_def_imped}
\delta_\ell  (r) = \sqrt{\frac{\varepsilon_0}{\mu_0}} \frac{E_\theta(r, \theta, \varphi)}{H_\varphi(r, \theta, \varphi)} \, .
\end{equation}
This approach is advantageous since the electromagnetic fields satisfy the same boundary conditions as in the massless case~\cite{Kroll1}. Using Eq.~\eqref{eq_def_U}, from Eqs.~\eqref{eq_ampere2} and~\eqref{eq_def_H_E} we find 
\begin{equation}
E_\theta = \frac{1}{r n^2}  \frac{\partial^2 ( \sqrt{n^2} r U ) }{\partial r \partial \theta}  \quad {\rm and} \quad
H_\varphi = \frac{i\omega\varepsilon_0}{r}  \frac{\partial ( \sqrt{n^2} r U ) }{ \partial \theta} \, ,
\end{equation}
which do not contain $m_\gamma$ explicitly and give
\begin{equation} \label{eq_def_imped2}
\delta_\ell  (r) = -\frac{i}{k (n^2)^{3/2} ru(r)} \frac{d ( \sqrt{n^2} r u_\ell (r) ) }{dr}  \, .
\end{equation}
Differentiating Eq.~\eqref{eq_def_imped2} and using Eq.~\eqref{eq_Ar_full}, we get
\begin{equation} \label{eq_diff_imped}
\frac{d\delta_\ell  (r)}{dr} + ik n^2 \delta_\ell^2 (r) - ik \left( 1 - \frac{\gamma^2}{k^2 n^2}  \right) = 0 \, ,
\end{equation}
where it is clear that the effects of a finite photon mass are suppressed in regions of high conductivity such as the upper atmosphere, cf. Fig.~\ref{fig_cond_profiles}, or Earth's interior.

Let us now work out the boundary conditions on $\delta_\ell  (r)$. The tangential components of the electric field are continuous across the boundary, irrespective of $m_\gamma$. The tangential components of the magnetic field, however, are discontinuous. Since the electromagnetic fields are zero inside Earth (cf. Sec.~\ref{sec_const_cond}), directly above the ground we have $E_\theta(r, \theta, \varphi) = 0$, whereas $H_\varphi(r, \theta, \varphi) \neq 0$. Therefore, we have that $\delta_\ell (R_\oplus) = 0$.

The upper boundary at $r = r_{\rm top}$ is an idealization, since the atmosphere is an unbounded medium. While positive and negative ions prevail at lower altitudes~\cite{Burke}, the lower layers of the ionosphere (D region) are dominated by free electrons, therefore characterizing a plasma with electron number density $n_e \approx 10^{10}-10^{12} / {\rm m}^3$~\cite{Hajj}. The plasma frequency is $\omega_p^2 = n_e e^2/m_e \varepsilon_0$, so that $f_p = \omega_p/2\pi  \sim \mathcal{O}\left( 1 \, {\rm MHz} \right)$, thus implying that ELF waves with $f_\ell \ll f_p$ are reflected. This effect is explored in the global transmission of long-wave radio signals.

The atmospheric layers above this top height will then have negligible effect on the results given some desired precision. Therefore, for $r \geq r_{\rm top}$ we have an effectively homogeneous ionosphere with $|n^2_{\rm top}| \gg 1$, a constant. At, say, $r_{\rm top} \approx R_\oplus + 90$~km and with $f = 10$~Hz we have $|n^2_{\rm top}| \approx 4 \times 10^4$, cf. Fig.~\ref{fig_cond_profiles}, and $|k|^2 \approx 4 \times 10^{-8} \, {\rm km}^{-2}$, whereas $\mu_\gamma^2 \approx 2 \times 10^{-9} \, {\rm km}^{-2}$, cf. Eq.~\eqref{eq_mu}. With these values, we have that $k^2 n^2_{\rm top} \gg \mu_\gamma^2$ and we may disregard the photon mass at the upper atmospheric layers. The boundary condition at $r = r_{\rm top}$ is then~\cite{Bliokh}
\begin{equation} \label{eq_boundary_iono}
\delta_\ell (r_{\rm top}) \approx 1/\sqrt{n^2_{\rm top}} \, .
\end{equation}

We now solve Eq.~\eqref{eq_diff_imped} from $r_{\rm top} = R_\oplus + z_{\rm top}$, where Eq.~\eqref{eq_boundary_iono} must be satisfied, until $r = R_\oplus$ with $\delta_\ell (R_\oplus) = 0$. Starting at some chosen $r_{\rm top}$, we may obtain the frequencies via Newton-Raphson's method
\begin{equation} \label{eq_approx_k}
k_\ell^{(j + 1)} = k_\ell^{(j)} - \frac{ \delta_\ell (R_\oplus, k_\ell^{(j)}) }{ \left. \frac{\partial }{\partial k } \left[ \delta_\ell (R_\oplus, k) \right] \right|_{k = k_\ell^{(j)}}  } \, 
\end{equation}
with $j = 0,1,2,...$ indexing the iteration step. Conveniently defining $\delta_{\ell, k} (r) \equiv \partial \delta_{\ell}(r)/\partial k$ and differentiating Eq.~\eqref{eq_diff_imped}, we find
\begin{equation} \label{eq_diff_imped_k}
\frac{d\delta_{\ell, k}  (r)}{dr} + 2ikn^2 \delta_\ell (r) \delta_{\ell, k} (r) + i \! \left[ \delta_\ell^2 (r) - \frac{\gamma^2}{\left( kn^2 \right)^2}  - 1\right] \! = \! 0,
\end{equation}
which must be solved concurrently with Eq.~\eqref{eq_diff_imped} subject to the boundary conditions $\delta_{\ell, k} (R_\oplus) = 0$ and
\begin{equation}
\delta_{\ell, k}  (r_{\rm top}) = \frac{n^2_{\rm top} - 1}{2k} \left[ n^2_{\rm top} \right]^{-3/2}  \, .
\end{equation}

The initial guesses for Eq.~\eqref{eq_approx_k} are taken from Eq.~\eqref{eq_SR_ideal} for a lossless cavity. Advancing with steps $\Delta z = 1$~km from the arbitrarily chosen top height $z_{\rm top} = 70$~km, we find that the frequencies are stable (within $< 1\%$) for a maximum height $z_{\rm top}^{\rm max} \geq 95$~km, consistent with the findings from Refs.~\cite{ACDC, Galuk_knee}. For $m_\gamma = 0$, profile~II~\cite{ACDC}, cf. Fig.~\ref{fig_cond_profiles}, offers the best match to measurements and will be adopted henceforth. Next, we include a finite photon mass by writing $\mu_\gamma = \left( 0.3/R_\oplus \right) \tilde{m}_\gamma$, cf.~Eq.~\eqref{eq_mu}, and sampling the range $\tilde{m}_\gamma = 0.1 - 7$ at steps of 0.1. For each mode we then determine which $\tilde{m}_\gamma$ causes a deviation from the measured frequency that saturates the estimated experimental uncertainties.

At this point it is worth noting that in Ref.~\cite{Kroll2} a fixed height of 70~km is assumed. Here, on the contrary, we iteratively find an effectively maximum height beyond which no improvement in the results is attained -- we could work with any height higher than this, but with no further benefit. Moreover, at $z = 70$~km, we have $\sigma \approx10^{-7}$~S/m, a factor $\sim 10^4$ lower than on the surface (or $\sim 10^7$ than on the oceans), cf. Fig.~\ref{fig_cond_profiles}. To be consistent with our assumption -- also made in Ref.~\cite{Kroll2} -- that Earth is a perfect conductor, the upper boundary of the cavity should be placed at a height $\gtrsim 90$~km.

Before we discuss the experimenntal uncertainties, let us briefly describe how Schumann-resonance data are typically taken and processed. Sensitive magnetometers~\cite{Salinas} or ball antennas~\cite{Modra_1, Modra_2, Ogawa} are set up to measure determined components of the ambient electric or magnetic fields. These fields represent the (incoherent) superposition of the effects from several sources at different locations worldwide at a frequency of $\sim$40 events per second~\cite{Oliver_lightning}, each with a different current spectrum modulating its amplitude. The raw data in the time domain are then Fourier transformed into the frequency domain, whereupon undesirable noise may be filtered from the resulting spectra ({\it e.g.}, noise from anthropogenic sources such as the electricity grid at~$50$~Hz).

The amplitude spectra display several broad peaks around the eigen-frequencies of the cavity, as expected on theoretical grounds~\cite{Sentman96}. The eigen-frequencies, as well as the respective quality factors, are then read from the positions and widths of the maxima of the distributions~\cite{LJones}. This task is typically accomplished via numerical fitting procedures ({\it e.g.}, by employing Lorentz-like~\cite{Modra_1, Modra_2, Salinas, Lorentz_fit} or Gaussian~\cite{gauss} fitting curves).

Let us now return to the estimation of the experimantal uncertainties. Monthly averaged daily variations of the fundamental mode are typically $\approx 0.5$~Hz~\cite{Modra_1, Modra_2, Satori}, though smaller changes in the range $0.04-0.14$~Hz have been reported during strong solar events~\cite{Schlegel}. For the second and third modes larger variations of respectively $\approx 1.0$~Hz and $\approx 1.2$~Hz may be inferred, particularly from Ref.~\cite{Modra_2}. We thus take $\delta f_\ell^{\rm exp} = \pm 0.25, \pm 0.5$ and $\pm0.6$~Hz for $\ell = 1,2,3$ as optimistic estimates for the uncertainties. Note that the word ``uncertainty" here refers not to the numerical error related to a certain data point (eigen-frequencies in a given spectra), but rather to the variability of the determined eigen-frequencies in the spectra obtained in different days, months, etc.

\begin{figure}[!t]
\begin{center}
\begin{minipage}[b]{1.12\linewidth}
\includegraphics[width=\linewidth]{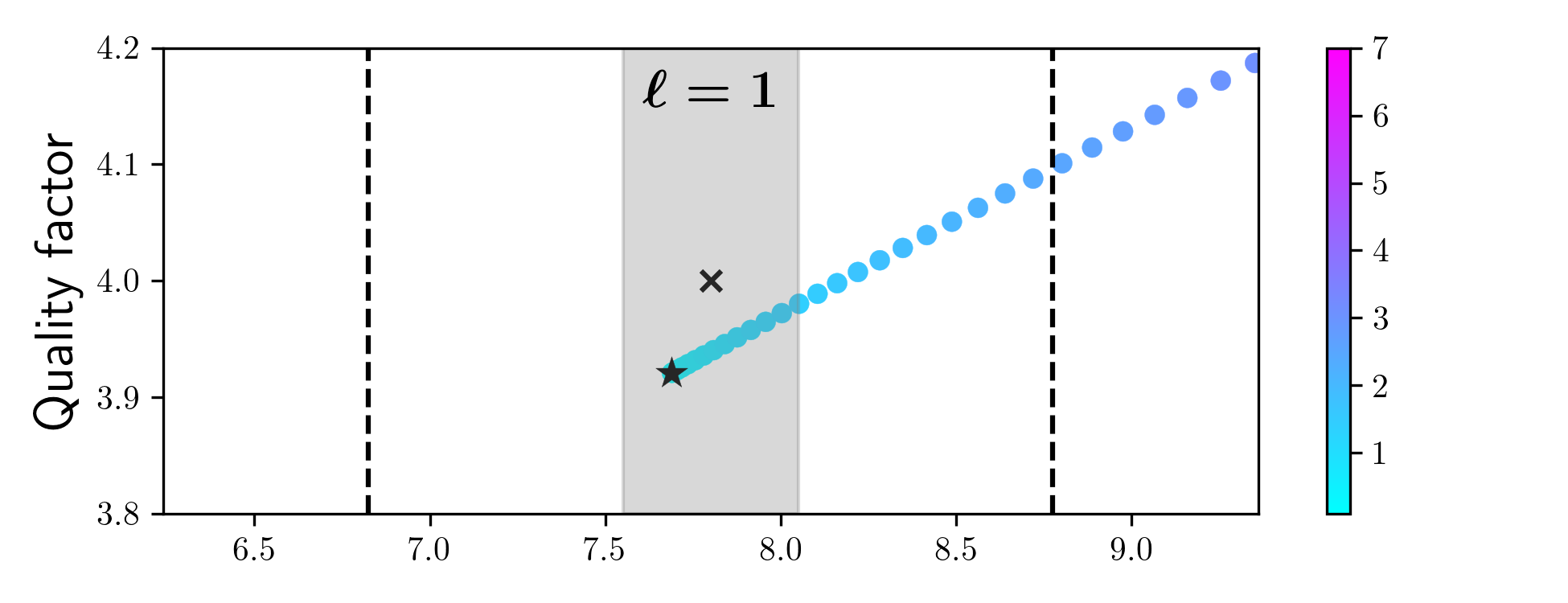}
\end{minipage} 
\begin{minipage}[b]{1.12\linewidth}
\includegraphics[width=\linewidth]{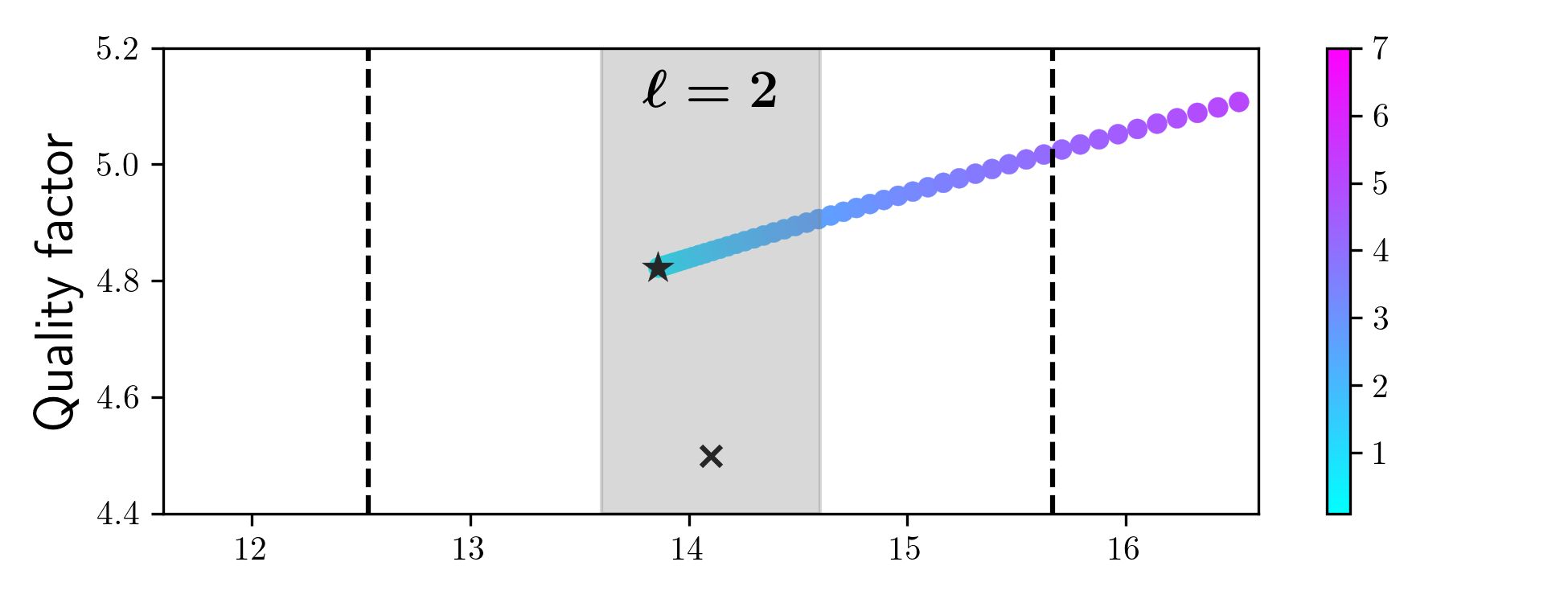}
\end{minipage}
\begin{minipage}[b]{1.12\linewidth}
\includegraphics[width=\linewidth]{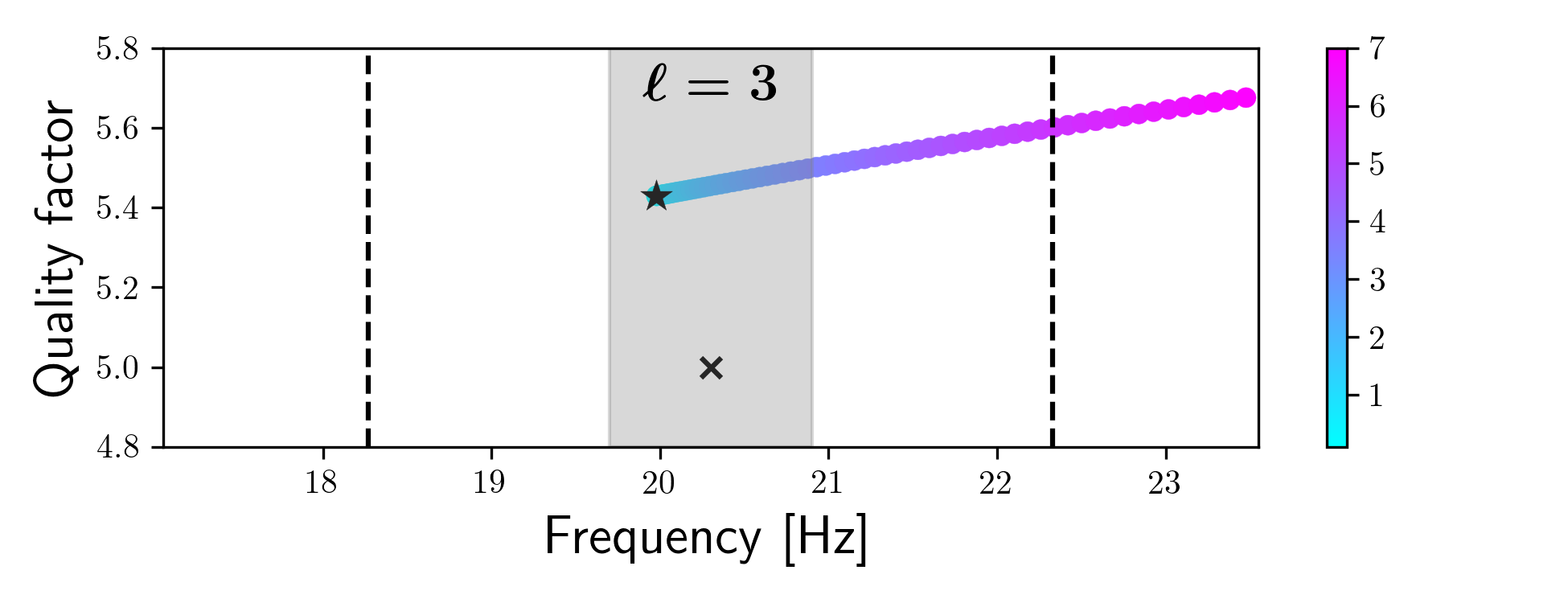}
\end{minipage}
\caption{General results for the first three modes ($\ell = 1,2,3$). Experimental data~\cite{Balser} and calculated pairs $(f_\ell, Q_\ell)$ obtained with profile~II~\cite{ACDC} are shown with black crosses and stars, respectively (both with $m_\gamma = 0$). The shaded regions correspond to $f_\ell^{\rm exp} \pm \delta f_\ell^{\rm exp}$, whereas the dashed lines indicate the regions within $f_\ell^{\rm exp} \pm \Delta f_\ell^{\rm exp}$ (see text for details). The colored points represent $(f_\ell, Q_\ell)$ assuming a finite photon mass with the color scale depicting $m_\gamma$ in units of $10^{-14} \, {\rm eV/c^2}$. }
\label{fig_f_Q_m}
\end{center}
\end{figure}

A second estimate may be obtained by noting that, due to the finite atmospheric conductivity, the $k_\ell$ obtained via Eq.~\eqref{eq_approx_k} are complex. The quality factor of a resonating cavity, defined as $Q_\ell = {\rm Re}(k_\ell) /2 \, {\rm Im}(k_\ell)$~\cite{Jackson}, is a measure of how lossy the cavity is -- a lossless cavity has real eigen-frequencies and therefore an infinite quality factor. The measured noise power spectra display spaced peaks with relatively broad widths. These spectra are typically fitted by Lorentzian curves with the eigen-frequencies being identified as the central peaks for each mode. For real cavities, the peaks in amplitude are not infinitely sharp and a full width at half maximum $\Delta f_\ell$ may be extracted from the data. Interestingly enough, the quality factor can also be expressed in terms of $\Delta f_\ell$ as $Q_\ell \approx f_\ell / \Delta f_\ell$, which in practice allows an indirect assessment of the quality factors~\cite{Jackson, Ogawa}. With the pairs $\left( f_\ell^{\rm exp}, Q_\ell^{\rm exp} \right)$ from Ref.~\cite{Balser}, we then set $\Delta f_\ell^{\rm exp} = \pm f_\ell^{\rm exp}/2Q_\ell^{\rm exp}$ as conservative estimates for the size of meaningfully measurable variations around the central frequencies.

Setting $z_{\rm top}^{\rm max} = 100$~km for definitiveness, typical runs of Newton-Raphson's procedure require $\sim 5$ iterations to reach relative differences below $10^{-7}$ in frequency. The results for $m_\gamma = 0$ (black stars) and $m_\gamma > 0$ (color scale) are shown in Fig.~\ref{fig_f_Q_m}. Clearly, finite photon masses tend to increase both frequencies and quality factors. We are then able to derive two upper bounds per mode: an optimistic ($m_\gamma^{\rm opt}$) and a conservative ($m_\gamma^{\rm con}$), corresponding to the calculated $(f_\ell, Q_\ell)$ pairs crossing the 1-$\sigma$ lines $f_\ell^{\rm exp} + \delta f_\ell^{\rm exp}$ and $f_\ell^{\rm exp} + \Delta f_\ell^{\rm exp}$, respectively. From Fig.~\ref{fig_f_Q_m} we see that the tightest limits come from the fundamental mode ($\ell = 1$) and read 
\begin{subequations}
\begin{eqnarray}
m_\gamma^{\rm opt} & \leq & 1.4 \times 10^{-14} \, {\rm eV/c}^2 \, , \label{eq_bound_opt} \\
m_\gamma^{\rm con} & \leq & 2.5 \times 10^{-14} \, {\rm eV/c}^2 \, . \label{eq_bound_cons} 
\end{eqnarray}
\end{subequations}
The latter represents an almost ten-fold improvement upon Kroll's earlier conservative assessment~\cite{Kroll2, PDG}.

%%%%%%%%%%%%%%%%%%%%%%%%%%%%%%%%%%%%%%%%%%

\section{Concluding remarks}  \label{sec_conclusions}
\indent

In this paper we discussed Schumann resonances in the context of a finite photon mass. In order to constrain $m_\gamma$, we considered realistic atmospheric conductivity profiles, determining the eigen-frequencies and quality factors of the Earth-ionosphere cavity. For $m_\gamma = 0$, these are in good agreement with data inferred from observations. We numerically determine the influence of a a finite photon mass on the eigen-frequencies and, upon comparison with experimental data, we are able to place the competitive bounds~\eqref{eq_bound_opt} and~\eqref{eq_bound_cons}, superseding the latest (reliable) estimate by this method~\cite{Goldhaber, Kroll2}.

Our direct approach to the bounds is based on the fact that the observed data are compatible with Maxwell's massless electrodynamics. We then assume that any contribution from new physics must be hidden within the uncertainties (defined in terms of the variability of the Schumann-resonance paramters as discussed in Sec.~\ref{sec_results}). A more involved analysis would require the inclusion of the photon mass in the calculation of the theoretical amplitude spectra ({\it e.g.}, following Ref.~\cite{Sentman96}) -- for this, one must explicitly take the sources and their distribution worldwide into account. The next step would be to compare the theoretical spectra with the processed (observed) spectra searching for the maxima (the eigen-frequencies), the bandwidths (related to the Q factors) and the maximum value of the photon mass compatible with the data. This task would require an in-depth re-analysis of the data-taking and -processing procedures and we do not expect a significant improvement on the bounds~\eqref{eq_bound_opt} and~\eqref{eq_bound_cons}.

Further qualitative improvements could be attained by including day-night asymmetries (mainly due to reduced ion production from the solar wind at night~\cite{Galuk, Rycroft}) and the geomagnetic field. Such a treatment is in principle possible in 2-D or 3-D via numerical techniques such as finite-difference time domain (FDTD) analysis~\cite{FDTD}.

As a final remark, we note that larger systems are more sensitive to smaller photon masses. It could thus be interesting to expand our present analysis to Schumann resonances in other planets of the solar system. Particularly relevant would be the gas giants, the largest of which is Jupiter with $R_{\rm Jup} \approx 69911$~km. Since $R_{\rm Jup}/R_\oplus \approx 11$, the Jovian eigen-frequencies are expected to be an order of magnitude lower than on Earth, but sensitive enough instruments placed on orbiters (such as those on board of the C/NOFS satellite~\cite{CNOFS}) could remotely detect Schumann spectra. From Eq.~\eqref{eq_mu} we may naively expect a sensitivity to photon masses around $m_\gamma \lesssim 10^{-15} \, {\rm eV/c^2}$, but a more realistic estimate would need to take into account several factors, specially concerning the theoretical modelling of the Jovian electromagnetic environment~\cite{Jupiter} and the uncertainties in the data from instruments on board of satellites.

%%%%%%%%%%%%%%%%%%%%%%%%%%%%%%%%%%%%%%%%%%%%%%%%%%%

\begin{acknowledgments}
The authors are thankful for the constructive criticism received from the referees. We are grateful to A.D.A.M. Spallicci, A.K. Kohara, C.A. Zarro, M.V. dos Santos, M.M. Candido and P. de Fabritiis for helpful comments. PCM is indebted to Marina and Karoline Selbach for insightful discussions.
\end{acknowledgments}


\begin{thebibliography}{99}

% Intro %%%%%%%%%%%%%%%%


%\bibitem{Einstein1} A. Einstein, {\it \"Uber einen die Erzeugung und Verwandlung des Lichtes betreffenden heuristischen Gesichttspunkt}, Annalen der Physik {\bf 6}, 132 (1905).

%\bibitem{Einstein2} A. Einstein, {\it Ist die Tr\"agheit eines K\"orpers von seinem Energieinhalt abh\"angig?}, Annalen der Physik {\bf 13}, 639 (1905).

\bibitem{PDG} R.I. Workman {\it et al.}, Particle Data Group, Progr. Theor. Exp. Phys. p. 083C01 (2022).

\bibitem{MHD} D.D. Ryutov, {\it Using plasma physics to weigh the photon}, Plasma Phys. Control. Fusion {\bf 49}, 429 (2007).

\bibitem{Bonetti} L. Bonetti {\it et al.}, {\it FRB 121102 casts new light on the photon mass}, Phys. Lett. B {\bf 768}, 326 (2017).

\bibitem{Bentum} M.J. Bentum, L. Bonetti, A.D.A.M. Spallicci, {\it Dispersion by pulsars, magnetars, fast radio bursts and massive electromagnetism at very low radio frequencies}, Adv. Space Res. {\bf 59}, 736 (2017).

\bibitem{Wang} H. Wang, X. Miao, L. Shao, {\it Bounding the photon mass with cosmological propagation of fast radio bursts}, Phys. Lett. B {\bf 820}, 136596 (2021).

\bibitem{Retino} A. Retin\`o, A.D.A.M. Spallicci, A. Vaivads, {\it Solar wind test of the de Broglie-Proca massive photon with Cluster multi-spacecraft data}, Astropart. Phys. {\bf 82}, 49 (2016).

\bibitem{Davis} L. Davis, Jr., A.S. Goldhaber, M.M. Nieto, {\it Limit on the photon mass deduced from Pioneer-10 observations of Jupiter's magnetic field}, Phys. Rev. Lett. {\bf 35}, 1402 (1975).

\bibitem{Williams} E. R. Williams, J.E. Faller, H.A. Hill, {\it New experimental test of Coulomb’s law: a laboratory upper limit
on the photon rest mass}, Phys. Rev. Lett. {\bf 26}, 721 (1971).

\bibitem{Tu} L.-C. Tu, J. Luo, G.T. Gilles, {\it The mass of the photon} , Rep. Prog. Phys. {\bf 68}, 77 (2004).

\bibitem{Nieto} A.S. Goldhaber, M.M. Nieto, {\it Terrestrial and extraterrestrial limits on the photon mass}, Rev. Mod. Phys. {\bf 43}, 277 (1971).

\bibitem{Okun} L.B. Okun, {\it Photon: history, mass, charge}, Acta Phys. Pol. B {\bf 37}, 565 (2006).

\bibitem{Goldhaber} A.S. Goldhaber, M.M. Nieto, {\it Photon and graviton mass limits}, Rev. Mod. Phys. {\bf 82}, 939 (2010).

\bibitem{Goldhaber2} A.S. Goldhaber, M.M. Nieto, {\it How to catch a photon and measure its mass}, Phys. Rev. Lett. {\bf 26}, 1390 (1971).

\bibitem{Fischbach} E. Fischbach, H. Kloor, R.A. Langel, A.T.Y. Lui, M. Peredo, {\it New geomagnetic limits on the photon mass and on long-range forces coexisting with electromagnetism}, Phys. Rev. Lett. {\bf 73}, 514 (1994).

\bibitem{Fuellekrug} M. F\"ullekrug, {\it Probing the Speed of Light with Radio Waves at Extremely Low Frequencies}, Phys. Rev. Lett. {\bf 93}, 043901 (2004).

\bibitem{Kroll1} N.M. Kroll, {\it Theoretical interpretation of a recent experimental investigation of the photon rest mass}, Phys. Rev. Lett. {\bf 26}, 1395 (1971).

\bibitem{Kroll2} N.M. Kroll, {\it Concentric spherical cavities and limits on the photon rest mass}, Phys. Rev. Lett. {\bf 27}, 340 (1971).

\bibitem{schumann_orig} W.O. Schumann, {\it On the free oscillations of a conducting sphere which is surrounded by an air layer and an ionosphere shell} (in German), Z. Naturforsch. {\bf 7A}, 149 (1952).

\bibitem{JacksonHistory} J.D. Jackson, {\it Examples of the zeroth theorem of the history of physics}, Am. J. Phys. {\bf 76}, 704 (2008).

\bibitem{Besser2007} B.P. Besser, {\it Synopsis of the historical development of Schumann resonances}, Radio Sci. {\bf 42}, RS2S02 (2007).

\bibitem{Balser} M. Balser, C.A. Wagner, {\it Observations of Earth-ionosphere cavity resonances}, Nature {\bf 188}, 638 (1960).

\bibitem{Jackson} J.D. Jackson, {\it Classical electrodynamics}, 3rd edition, New York: John Wiley \& Sons. ISBN 978-0-471-30932-1. 

\bibitem{Pfaff} F. Sim\~oes, R. Pfaff, J.-J. Berthelier, J. Klenzing, {\it A review of low frequency electromagnetic wave phenomena related to tropospheric-ionospheric coupling mechanisms}, Space Sci. Rev. {\bf 168}, 551 (2012).

\bibitem{Williams1992} E.R. Williams, {\it The Schumann resonance: a global tropical thermometer}, Science {\bf 256}, 1184 (1992).

\bibitem{Hobara} M. Sekiguchi, M. Hayakawa, A.P. Nickolaenko, Y. Hobara. {\it Evidence on a link between the intensity of Schumann resonance and global surface temperature}, Ann. Geophys. {\bf 24}, 1809 (2006).

\bibitem{Price2000} C. Price, {\it Evidences for a link between global lightning activity and upper tropospheric water vapour}, Lett. Nature {\bf 406}, 290 (2000).

\bibitem{quakes} M. Hayakawa {\it et al.}, {\it Anomalous ELF phenomena in the Schumann resonance band as observed at Moshiri (Japan) in possible association with an earthquake in Taiwan}, Nat. Hazards Earth Syst. Sci. {\bf 8}, 1309 (2008).

\bibitem{Madden} T. Madden, W. Thompson, {\it Low-frequency electromagnetic oscillations of the Earth-ionosphere cavity}, Rev. Geophys. {\bf 3}, 211 (1965).

\bibitem{health_0} H. K\"onig, F. Ankerm\"uller, {\it \"Uber den Einfluss besonders niederfrequenter elektrischer Vorg\"ange in der Atmosph\"are auf den Menschen}, Naturwissenschaften {\bf 47}, 486 (1960).

\bibitem{health_1} S. J. Palmer, M. J. Rycroft, M. Cermak, {\it Solar and geomagnetic activity, extremely low frequency magnetic and electric fields and human health at the Earth’s surface}, Surv. Geophys. {\bf 27}, 557 (2006).

\bibitem{health_2} N.G. Ptitsyna, G. Villoresi, L.I. Dorman, N. Iucci, M.I. Tyasto, {\it Natural and man-made low-frequency magnetic fields as a potential health hazard}, Phys. Usp. {\bf 41}, 687 (1998).

%\bibitem{health_3} M. Destefanis {\it et al.}, {\it Extremely low frequency electromagnetic fields affect proliferation and mitochondrial activity of human cancer cell lines}, International Journal of Radiation Biology {\bf 91}, 1 (2015).

%\bibitem{Franken} P.A. Franken,  G.W, Ampulski, {\it Photon rest mass}, Phys. Rev. Lett. {\bf 2}, 115 (1971).

%\bibitem{Mewes} M. Mewes, {\it Bounds on Lorentz and CPT violation from the Earth-Ionosphere cavity}, Phys. Rev. D {\bf 78}, 096008 (2008).



% Basic theory %%%%%%%%%%%

\bibitem{dB1} L. de Broglie, {\it Radiations. -- Ondes et quanta}, Comptes Rendus Hebd. S\'eances Acad. Sc. Paris {\bf 177}, 507 (1923).

\bibitem{dB2} L. de Broglie, {\it Nouvelles Recherches sur la Lumi\`ere}, vol. 411 of Actualit\'es Scientifiques et Industrielles (Hermann \& Cie, Paris, 1936).

\bibitem{dB3} L. de Broglie, {\it La m\'echanique ondulatoire du photon, Une novelle théorie de la lumi\`ere}, Hermann, Paris, 1940.

\bibitem{Proca1} A. Proca, {\it Sur l’equation de Dirac}, Compt. Rend. {\bf 190}, 1377 (1930).

\bibitem{Proca2} A. Proca, {\it Particules libres photons et particules « charge pure »}, J. Phys. Radium {\bf 8}, 23 (1937).

\bibitem{Sentman} D.D. Sentman, {\it Approximate Schumann resonance parameters for a two-scale height ionosphere}, J. Atmos. Terr. Phys. {\bf 52}, 35 (1990).

\bibitem{Oliver_lightning} J.E. Oliver, {\it Encyclopedia of World Climatology}, National Oceanic and Atmospheric Administration. ISBN 978-1-4020-3264-6 (2005). 

\bibitem{Kakona} J. K\'akona {\it et al.}, {\it In situ ground-based mobile measurement of lightning events above central Europe}, EGUsphere [preprint], https://doi.org/10.5194/egusphere-2022-379 (2022).

\bibitem{Lopez} J.A. Lop\'ez {\it et al.}, {\it Spatio-temporal dimension of lightning flashes based on three-dimensional Lightning Mapping Array}, Atmospheric research {\bf 197}, 255 (2017).

\bibitem{Sentman96} D.D. Sentman, {\it Schumann resonance spectra in a two-scale-height Earth-ionosphere cavity}, J. Geophys. Res. Planets {\bf 101}, 9479 (1996).

%\bibitem{Sommerfeld} A. Sommerfeld, {\it Partial Differential Equations in Physics}, Academic Press, New York, 1949.






% r < R

\bibitem{world_atlas_cond} World Atlas of Ground Conductivities, Rec. ITU-R P.832-4, International Telecommunication Union, 2015.

\bibitem{Maus} S. Maus, {\it Electromagnetic ocean effects}, tech. rep., National Geophysical Data Center, NOAA E/GC1, 325 Broadway, Boulder, CO 80305-3328, 2003.

\bibitem{Rycroft} M.J. Rycroft, {\it Some effects in the middle atmosphere due to lightning}, J. Atmos. Terr. Phhys {\bf 56}, 343 (1994).
\bibitem{Price1} A.T. Price, {\it The electrical conductivity of the Earth}, Q. J. R. Astron. Soc. {\bf 11}, 23 (1970).

\bibitem{Poirier} J. Peyronneau, J.P. Poirier, {\it Electrical conductivity of the Earth’s lower mantle}, Nature {\bf 342}, 537 (1989).

\bibitem{mantle2} V.R.S. Hutton, {\it The electrical conductivity of the Earth and planets}, Rep. Prog. Phys. {\bf 39}, 487 (1976).

\bibitem{cole} R.K. Cole, Jr., {\it The Schumann resonances},  Radio Sci. {\bf 69}, 1345 (1965).

%\bibitem{mantle3} M. Pozzo, C. Davies, D. Gubbins, D. Alf\`e, {\it Thermal and electrical conductivity of iron at Earth's core
%conditions}, Nature {\bf 485}, 355 (2012).


% r > R

\bibitem{ACDC} I.G. Kudintseva, A. P. Nickolaenko, M.J. Rycroft, A. Odzimek, {\it AC and DC global electric circuit properties and the height profile of atmospheric conductivity}, Annals of Geophysics {\bf 59}, 5 (2016).

\bibitem{Burke} R. Sagalyn, H. Burke, {\it Atmospheric Electricity, Handbook of Geophysics and the Space Environment}, (A. S. Jursa, ed.), ch. 20.1. Air Force Geophysics Laboratory, Air Force Systems Command, United States Air Force, 1985.

\bibitem{Greifinger} C. Greifinger, P. Greifinger, {\it Approximate method for determining ELF eigenvalues in the earth-ionosphere waveguide}, Radio Sci. {\bf 13}, 831 (1978).

\bibitem{cond_prof_galuk} A.P. Nickolaenko, Yu.P. Galuk, M. Hayakawa  {\it Vertical profile of atmospheric conductivity that matches Schumann resonance observations}, SpringerPlus {\bf 5}, 108 (2016).

\bibitem{Bliokh} P.V. Bliokh, A.P. Nicholaenko, Yu. F. Filippov, {\it Schumann resonances in the Earth-ionosphere cavity}, Inst. of Elec. Eng. Waves Ser., vol. 9, Peter Peregrinus, London.

\bibitem{Titan} C. B\'eghin {\it et al.}, {\it Analytic theory of Titan’s Schumann resonance: Constraints on ionospheric conductivity and buried water ocean}, Icarus {\bf 218}, 1028 (2012).

\bibitem{Arfken} G. Arfken, H. Weber, F. Harris, {\it Mathematical Methods for Physicists}, Academic Press; 7th ed..

\bibitem{Hamelin} F. Sim\~oes {\it et al.}, {\it Electromagnetic wave propagation in the surface-ionosphere cavity of Venus}, J. Geophys. Res. Planets {\bf 113}, E07007 (2008).






% Conductivity profiles



%\bibitem{Mushtak2002} V.C. Mushtak, E. Williams, {\it ELF propagation parameters for uniform models of the Earth-ionosphere waveguide}, J. Atmos. Sol. Terr. Phys., {\bf 64}, 1989 (2002).

%\bibitem{cummer} S.A. Cummer, {\it Modeling electromagnetic propagation in the Earth-ionosphere waveguide}, IEEE Trans. Antennas Propag. {\bf 48}, 1420 (2000).

%\bibitem{Ampferer} M. Ampferer, V.V. Denisenko, W. Hausleitner, S. Krauss, G. Stangl, M.Y. Boudjada, H.K. Biernat, {\it Decrease of the electric field penetration into the ionosphere due to low conductivity at the near ground atmospheric layer}, Ann. Geophys. {\bf 28}, 779 (2010).






% Results %%%%%%%%%%%%%%

\bibitem{Wait1970} J.R. Wait, {\it Electromagnetic Waves in Stratified Media}, 2nd edition, Pergamon Press, Oxford, 608 p. (1970).

\bibitem{Hajj} G.A. Hajj, L.J. Romans, {\it Ionospheric electron density profiles obtained with the Global Positioning System: Results from the GPS/MET experiment},  Radio Sci. {\bf 33}, 175 (1998).

\bibitem{Galuk_knee} Yu.P. Galuk, A.P. Nickolaenko, M. Hayakawa, {\it Knee model: Comparison between heuristic and rigorous solutions for the Schumann resonance problem}, J. Atmos. Sol.-Terr. Phys. {\bf 135}, 1364 (2015).

\bibitem{Salinas} A. Salinas {\it et al.}, {\it Schumann resonance data processing programs and four-year measurements from Sierra Nevada ELF station}, Computers \& Geosciences {\bf 165}, 105148 (2022).

\bibitem{Modra_1} A. Ondraskova, S. Sevcic, P. Kostecky, {\it A significant decrease of the fundamental Schumann resonance frequency during the solar cycle minimum of 2008-9 as observed at Modra Observatory}, Contributions to Geophysics and Geodesy {\bf 39/4}, 345 (2009).

\bibitem{Modra_2} A. Ondraskova, S. Sevcic, P. Kostecky, L. Rosenberg, {\it Long-term observations of Schumann resonances at Modra Observatory}, Radio Sci. {\bf 42}, RS2S09 (2007).

\bibitem{Ogawa} T. Ogawa, Y. Tanaka, {\it Q factors of the Schumann resonances and solar activity}, Special Contributions of the Geophysical Institute, Kyoto University {\bf 10}, 21 (1970).

\bibitem{LJones} D.L. Jones, {\it  Schumann Resonances and E.L.F. propagation for inhomogeneous, isotropic ionosphere profles}, Journal of Atmospheric and Terrestrial Physics {\bf 29}, 1037 (1965).

\bibitem{Lorentz_fit} V.C. Mushtak, E.R. Williams, {\it An improved Lorentzian technique for evaluating resonance characteristics of the Earth-ionosphere cavity}. Atmospheric Research {\bf 91}, 188 (2009).

\bibitem{gauss} J. Rodriguez-Camacho {\it et al.}, {\it On the need of a unified methodology for processing Schumann resonance measurements}, J. Geophys. Res. Atmos. {\bf 123}, 13277 (2018).

\bibitem{Satori} G. Satori, {\it Monitoring Schumann resonances -- II. Daily and seasonal frequency variations}, J. Atmos. Terr. Phys. {\bf 58}, 1483 (1996).

\bibitem{Schlegel} K. Schlegel, M. F\"ullekrug, {\it Schumann resonance parameter changes during high-energy particle precipitation}, J. Geophys. Res. {\bf 104}, 10,111 (1999).




% Conclusions %%%%%%%%%%

\bibitem{Galuk} Yu.P. Galuk, A.P. Nickolaenko, M. Hayakawa, {\it Impact of the ionospheric day-night non-uniformity on the ELF radio-wave propagation}, Radiophys. Quantum El. {\bf 61}, 176 (2018).

\bibitem{FDTD} H.H. Yang, V.P. Pasko, {\it Three-dimensional finite difference time domain modeling of the Earth-ionosphere cavity resonances}, Geophhys. Res. Lett. {\bf 32}, L03114 (2005).

\bibitem{CNOFS} F. Sim\~oes, R.F. Pfaff, H. Freudenreich, {\it Satellite observations of Schumann resonances in the Earth's ionosphere}, Geophys. Res. Lett. {\bf  38}, L22101 (2011).

\bibitem{Jupiter} F. Sim\~oes {\it et al.}, {\it Using Schumann resonance measurements for constraining the water abundance on the giant planets -- implications for the solar system's formation}, The Astrophysical Journal {\bf 750}, 85 (2012).

% Appendix %%%%%%%%%%










\end{thebibliography}
\end{document}